\documentclass[english,aps,pra,reprint,noshowpacs,superscriptaddress,nofootinbib]{revtex4-1}   
\usepackage[T1]{fontenc}	
\usepackage[latin9]{inputenc}	
\usepackage{geometry}		
\usepackage{graphicx}
\usepackage[above,below]{placeins}	
\usepackage{times}

\usepackage{hyperref}
\usepackage{array}
\hypersetup{colorlinks=true,urlcolor=blue,citecolor=blue,linkcolor=blue}   
\begin{document}

\title{Computational Thinking in Introductory Physics}
\author{Orban, C. M.}
\affiliation{Physics Department, The Ohio State University, 191 W. Woodruff Ave., Columbus, Ohio, 43210}
\author{Teeling-Smith, R. M.}
\affiliation{University of Mt. Union, Alliance, OH, 44601}

\maketitle

\section{Introduction}

\label{sec_intro} ``Computational thinking" (CT) is still a relatively new term in the lexicon of learning objectives and science standards. The term was popularized in an essay by Wing \cite{Wing2006} who said ``To reading, writing and arithmetic, we should add computational thinking to every child's analytical ability". Agreeing with this premise, in 2013 the authors of the Next Generation Science Standards (NGSS) included ``mathematical and computational thinking" as one of eight essential science and engineering practices that K-12 teachers should strive to develop in their students \cite{ngss}. 

There is not yet widespread agreement on the precise definition or implementation of CT, and efforts to assess CT are still maturing, even as more states adopt K-12 computer science standards \cite{k12CSstandards}. In this article we will try to summarize what CT means for a typical introductory (i.e. high school or early college) physics class. This will include a discussion of the ways that instructors may already be incorporating elements of CT in their classes without knowing it. 

Our intention in writing this article is to provide a helpful introduction to this topic for physics instructors, which is a very different goal than providing a rigorous survey of the literature. For more rigor, interested readers should consult Weintrop et al. \cite{Weintrop_etal2016}, Sengupta et al. \cite{Sengupta_etal2013}, and Grover \& Pea~\cite{Grover2013}.

We hope that our comments here will also be useful to the growing number of physics instructors who are integrating computer science (CS) into their classrooms through coding activities in VPython \cite{Chabay_Sherwood2008}, JavaScript and other languages. Groups like PICUP and AAPT have a number of resources and workshops to facilitate this work \cite{aaptComp,picup}. We ourselves lead an effort called the STEMcoding project which focuses on coding activities for high school and early college physics \cite{Orban_etal2018}.

For brevity, in this article we will not discuss ``unplugged" CT activities even though CT does not always require a computer, which is perhaps the first thing to appreciate about it. This is because humans compute too!  

\section{Defining Computational Thinking}
\label{sec:define}

One of the most highly cited papers on CT that is also relatively recent is Weintrop et al. \cite{Weintrop_etal2016}. Having noticed that experts were defining CT in different ways, Weintrop et al. went about identifying CT ``practices" from the literature, from sample activities they collected, and from interviewing both teachers and STEM professionals on what kind of skills they associate with CT. Over half of the activities they sampled were on the subject of physics, and their interviews included physics teachers and physics professionals. 

The summary in their Fig.~2 concludes that there are four main CT practices, each with similar importance: 
\begin{enumerate}
    \item {\bf Data Practices} -- Collecting, Creating, Manipulating, Analyzing and Visualizing Data
    \item  {\bf Modeling and Simulation Practices } -- Using Computational Models to Understand a Concept, Using Computational Models to Find and Test Solutions; Assessing, Designing, and Constructing Computational Models
    \item {\bf Computational Problem Solving Practices} -- Preparing Problems for Computational Solutions, Programming, Choosing Effective Computational Tools, Assessing different Approaches/Solutions to a Problem, Developing Modular Computational Solutions, Creating Computational Abstractions, Troubleshooting and Debugging 
    \item {\bf Systems Thinking } --  Investing a Complex System as a Whole, Understanding the Relationships within a System, Thinking in Levels, Communicating Information about a System, Defining Systems and Managing Complexity
\end{enumerate}

Only one of the four CT practices -- Computational Problem Solving -- is what one might typically associate with coding, programming, or debugging. In this way, the CT practices of \cite{Weintrop_etal2016} reflect the idea that computational thinking does not necessarily require a computer.

Physics instructors looking over this list will recognize ``Modeling and Simulation Practices" as familiar practices they likely already use in their classrooms. For some time now, simulations have been an important part of physics instruction through resources like PhET Interactive Simulations and Physlet Physics, as the paper acknowledges. Here the term ``modeling" is used in much the same way that it is in the physics education community. Weintrop et al.~\cite{Weintrop_etal2016} cites the NGSS, which was heavily influenced by the modeling movement and ``modeling instruction"  \cite{ngssPractices}, as a primary reference for this category. The connection between CT and modeling is an important philosophical foundation for the ``Bootstrap for physics" approach to integrating computer science into physics and physical science classrooms \cite{bootstrapphysics}, which is a curriculum developed by the American Modeling Teachers Association.

Another CT practice is ``Data Practices". A more self-explanatory title for ``Data Practices" might be ``Working with Data". The list of activities associated with this practice has a great deal of overlap with the goals of a typical physics lab activity. This raises the possibility that CT may contribute something useful to the current debate over the usefulness of physics labs (\cite{Homes_Wieman2018,Homes_Smith2019}). We will return to this topic later.



The last CT practice is ``Systems Thinking", which is the most abstract of the four.  For physics instructors, this practice can perhaps be understood as the skill of gaining insight from simulations designed to model situations that include one or more complex interactions. Computation is valuable in physics because it allows the ideas of physics to be usefully applied to problems that are difficult to treat analytically. A classic example of this is numerically modeling air drag on a projectile or falling object. Although the only forces are air drag and gravity, from a student's perspective, this is a complex simulation with competing effects. A possible ``systems thinking" activity would be to analyze the simulation to measure the terminal velocity and check if it agrees with the expectation from assuming balanced forces of drag and gravity. So even this, the most abstract CT practice, seems to fit naturally within the learning objectives of a typical physics course.


\section{What can Computational Thinking offer to Physics Instruction?}


The previous section outlines and clarifies the ways that physics instruction aligns with CT as characterized by Weintrop et al.~\cite{Weintrop_etal2016}. In this section we examine what computation / CT\footnote{Although there are the subtle differences between computation (or coding) and CT, as discussed in Sec.~\ref{sec:define}, for simplicity of language in this section we treat ``computation / CT" as one thing unless noted otherwise.} has to offer to physics instruction. Let us re-phrase this question slightly by asking: what are we trying to do when we integrate computational or algorithmic ideas in some form in an introductory physics class? How does this help the mission of an introductory physics class?

We can think of at least four types of responses to this question, which we outline below. The first two items have surely been echoed by many others, 
 while the last two perhaps less so.

\begin{enumerate}
    \item Computation / CT is valuable in introductory physics because it is an integral part of physics as a discipline.  Not including it would be a false characterization of what physics is, and/or would leave our students ill prepared for their future careers in various fields inside and outside of physics.
    \item Computation / CT is valuable in physics because setting up codes and commercial programs to accurately simulate a physical system is often a non-trivial task. We can begin to develop in our students an understanding of the subtleties of this task starting at the introductory level.
    \item Creating and debugging a program in order to model some physical phenomena naturally involves moments where one's intuition or understanding is puzzled or challenged in some way. The process of overcoming an intellectual challenge like this is generally called ``sense making" and it is among the most important things that happen in a physics classroom. Computation / CT is valuable because it provides an interesting new venue for ``sense making" in physics.
    \item Computation / CT is valuable as another ``representation" in physics. In addition to seeing physics through interdependent mathematical expressions (e.g. the equation sheet) or diagrams/graphs, students can see physics as iterative relationships in some kind of code.    
\end{enumerate}
In reading through this list, which is by no means exhaustive, it is important to appreciate that in an introductory physics class, relatively few of our students are on a physics or computer science career path. {\bf Introducing computation / CT into high school or
college non-major physics courses, in our opinion, should not be done solely to provide an early start to the kind of skills students would need as a professional physicist or as a software engineer.}
More often, we are teaching students who are heading into traditional technical or engineering careers. These students may never need advanced programming skills but many will go on to use sophisticated simulations to model any number of phenomena (e.g. stress and strain, heat flow, traffic patterns). For these students, introducing computation / CT into an introductory physics course provides an introduction to the skill of making sense of what is to them a sophisticated simulation. 

\begin{figure*}
    \centering
    \includegraphics[width=4in]{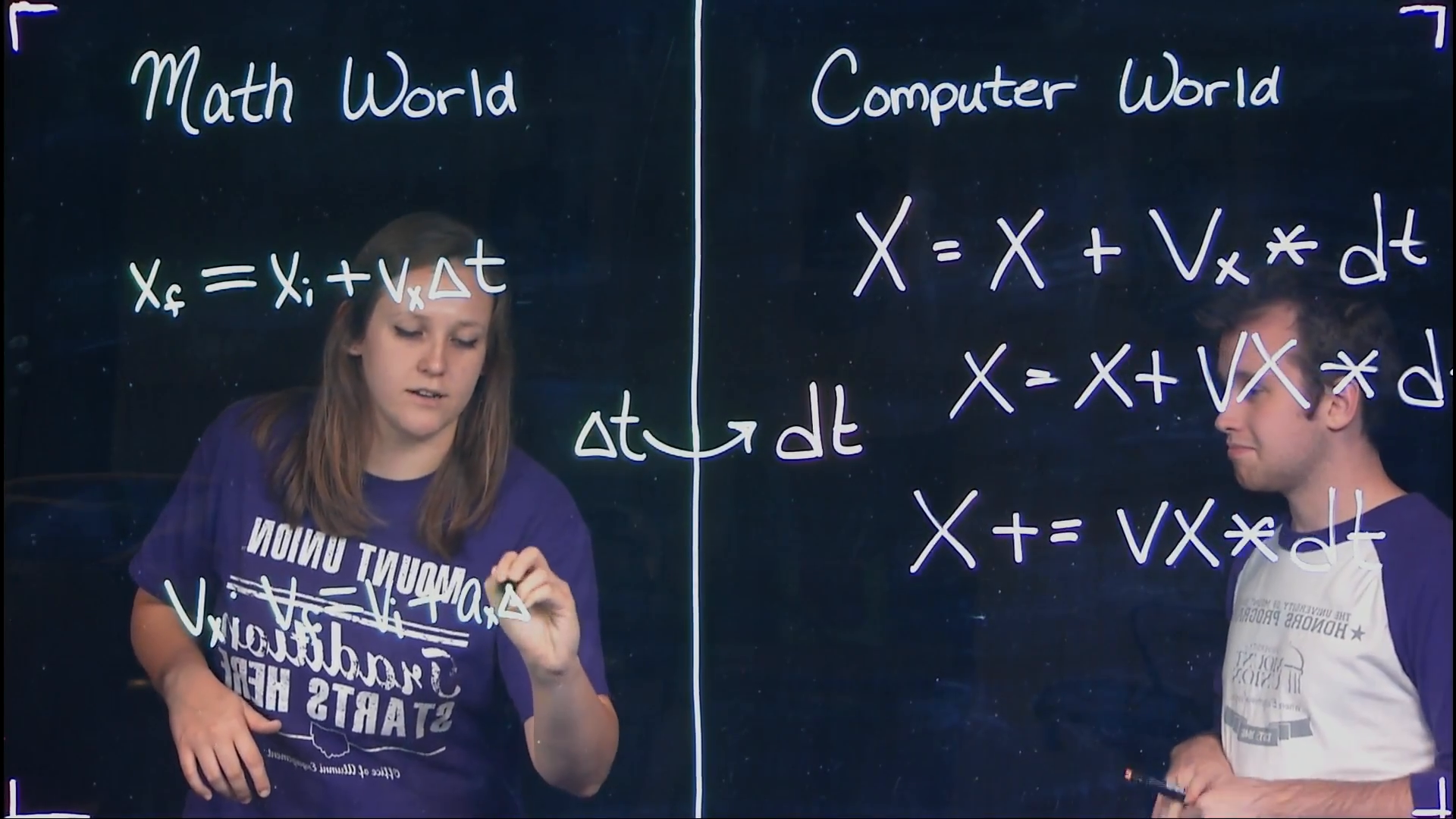}
    \caption{A screenshot from ``The Physics of Video Games" hour of code activity from the STEMcoding project, showing students discussing the differences between writing an expression in ``Math World" vs. in ``Computer World" \cite{HofCSTEMcoding}. We argue that understanding the differences between mathematical and computational representations is central to teaching computational thinking in the classroom and key to assessing this hard-to-define skill.}
    \label{fig:math_vs_cs}
\end{figure*}

This is an important backdrop to unpacking the statements on this list. The first statement, for example, talks about computation as a part of physics as a discipline \cite{Landau2006,PICUPrec}. More than the equations, it is \emph{the discipline} of physics that is relevant to engineering and other fields, which is why students may be ``ill prepared for their future careers in various fields" if computation / CT were ignored.

The second statement very much reflects the role that physical simulations play in engineering work.
A concise summary of the second statement would be helping students to understand the difference between ``math world" and ``computer world". As depicted in Fig.~\ref{fig:math_vs_cs}, students need to understand how to translate a mathematical expression to a computer code and that running a simulation often involves a finite time step. Both of these topics are important introductions to the kinds of sophisticated simulations/programs that students may encounter and use in the workforce.

Understanding the assumptions of a sophisticated program is key to better making sense of simulation results.  In this way, the third statement on ``sense making" is likewise important for engineering. 
Sense making is summarized by Tor Odden and collaborators as ``a dynamic process of building an explanation in order to resolve a gap or inconsistency in [one's] knowledge" \cite{Odden2017}.  It is not difficult to see the connection between sense making and CT if one considers the iterative process that students experience in debugging and running a code. One STEMcoding activity we have our students complete involves an asteroids-like game wherein we provide a simple 1D code for a ship traveling through free space that students modify into a 2D code.  Students reliably make mistakes, such as forgetting to initialize variables or copying the code for horizontal motion without making appropriate modifications for vertical motion \cite{Orban_etal2018}.  These mistakes produce perplexing behavior in the game and every step of identifying and correcting these errors engages the sense making skills of the student in addition to their computational skills.
A recent study where the connection between computation / CT and sense making is explicit is ``How computation can facilitate sensemaking in physics: a case study" by Petter-Sand, Odden et al. \cite{PetterSand_etal2018}. They identify sense making moments from interviews of students as they complete a coding activity on radioactive decay.

Sense making in this context can also include showing students how to check the accuracy of simulation against an analytic solution. {\bf We cannot let students assume that complex simulations are right simply because of their complexity.} With this as an objective, an emphasis is placed on code verification tasks and the student's conceptual physics knowledge becomes an important tool for for critical thinking and sense making about the program's result. 

Comparing simulation results to analytical expressions also reinforces the idea that the computer program is another ``representation" of the interaction, as discussed in the fourth statement. Using multiple representations is a key element of the widely influential modeling instruction approach \cite{Wells_etal1995}. The computational curriculum developed by the American Modeling Teachers Association naturally integrates computation as another representation \cite{bootstrapvideo}. This idea is also championed by well-known authors like Papert \cite{Papert1980}, diSessa \cite{Disessa2018} and their collaborators (e.g. \cite{Sherin2001}). Certainly this is part of the rationale for including ``modeling and simulation" as one of the key practice of CT in Weintrop et al. \cite{Weintrop_etal2016}, as mentioned earlier.

\section{Concrete Examples to Assess Computational Thinking}

In this section we outline two concrete examples of assessing computational thinking that involve students thinking critically about a section of code. In other CT assessments that we have seen (e.g. \cite{ctstem}), it is not unusual to show students a code and ask them to answer questions about it.

\subsection{Perfectly Inelastic Collisions}

In the ``Planetoids with Momentum!" activity from the STEMcoding project \cite{momentum} students take an asteroids-like code and modify it until the ship can collide and stick to a circle which is like a blob of goo drifting through space. In this way the activity illustrates a 2D perfectly inelastic collision. 

There are detailed directions for this activity that include the correct code to determine the final velocity of the ship and ``blob" which in 1D looks like this:
\begin{verbatim}
if (collided == true) { 
vx1 = (mass1*vx1 + mass2*vx2)/(mass1 + mass2) 
vx2 = vx1
}
\end{verbatim}

Students measure this final velocity and check that it matches with expectation from momentum conservation. Towards the end of this exercise, students are asked the following question: \\

\noindent \emph{The following code will give the wrong answer for the final velocity of the ship and blob after the collision:}
\begin{verbatim}
if (collided == true) { 
vx1 = (mass1*vx1 + mass2*vx2)/(mass1 + mass2) 
vx2 = (mass1*vx1 + mass2*vx2)/(mass1 + mass2)
}
\end{verbatim}
\emph{Copy this into your code, take out the expression you used before and run the code to see what happens. Why does this give a different (wrong) answer for the velocity after the collision?}

This is an interesting example because the code looks essentially identical to the mathematical solution for two objects colliding and sticking together in a 1D perfectly elastic collision. But when the code is run (as one can do at this link \url{http://go.osu.edu/momentumdemo}), the student finds that instead of near-perfect agreement with the expectation from momentum conservation, the program may be off from the correct final velocity by tens of percent. To understand the reason for this, students must appreciate that \texttt{vx1} is being updated and then the updated value of \texttt{vx1} is used again in the following line of code. This is wrong because the mathematical expression only uses the velocity from before the impact to compute the velocity after.

This example illustrates a key difference between mathematical and computational representations is that the computer goes line-by-line through the program whereas there is really no equivalent to this in ``math world" (i.e. high school or early college algebra). In general, students do not automatically look at a code and realize that the computer goes line-by-line or that the same code is run over and over again. Perhaps the first study of CT in introductory physics was by Aiken~et~al.~\cite{Aiken_etal2013} who found that high school students struggle to understand the iterative nature of computer programs.

The assessment just described also connects with ``sense making" and code verification as discussed earlier. Importantly, it shows that even what appears to be a correct implementation of an equation needs to be verified for accuracy. 

\subsection{Projectile Motion}

Projectile motion is important both for introductory physics and for introducing computation. In their paper ``Integrating Numerical Computation into the Modeling Instruction Curriculum"  Caballero et al. \cite{Caballero_etal2014} focuses on a VPython implementation of projectile motion. A key part of this code is the section that advances the velocity and position of the object. Typically this section would be written something like this:
\begin{verbatim}
// acceleration
ax = 0
ay = -9.8 

// update velocity
vx += ax*dt
vy += ay*dt

// update position
x += vx*dt
y += vy*dt

// time elapsed
t += dt;
\end{verbatim}
where \texttt{dt} is the timestep. Typically this code would be run over and over until some desired time. The code just described and the VPython code in \cite{Caballero_etal2014} are examples of Euler-Cromer integration \cite{Cromer1981}. This is a perfectly adequate way of representing projectile motion, but it is still an approximation because the position is updated after the velocity update, meaning that we are using the velocity at the \emph{end} of each timestep to advance the position instead of correctly using the average velocity \emph{during} the timestep. After many timesteps, the simulated position will stray from the analytic result ($y (t) = y_i + v_{xi}t + \frac{1}{2}a_y t^2$) and a careful analysis will show (as one can easily demonstrate from javascript code we have linked here: \url{http://go.osu.edu/energydemo}) that the kinetic plus potential energy in the simulation slowly decreases instead of staying constant as it should \cite{Cromer1981, roos}.

An interesting question to ask introductory physics students is \emph{what angle gives the farthest distance traveled for the projectile in the simulation?} 
Many students will hear this question and quickly answer that the code gives 45 degrees as the farthest distance. But students that carefully investigate will find that angles that are slightly off from 45 degrees allow the projectile to travel slightly farther (due to numerical inaccuracies). In this way, the question helps to identify how many of our students approach the simulation with a healthy skepticism. Instructors can also use this question as a launching point for a discussion about numerical methods.

Questions like these are reminiscent of the shift in thinking about physics labs (e.g. \cite{Homes_Wieman2018}). Instead of simply verifying the equations on the equation sheet, careful measurements can reveal when these equations fail. For example, Holmes \& Bonn 2015 \cite{Holmes_Bonn2015} describe an experiment where students discover that the period of a pendulum \emph{does} have a weak dependence on the amplitude. There is now some evidence that experiments like this that focus on critical thinking skills can be more impactful than focusing on concepts \cite{Wilcox2017}. Computational thinking can be part of this shift in emphasis, especially since computational activities in are often completed during lab sections.

\section{Thoughts for the Future}

Perhaps the most meaningful probe of CT is whether students can configure a code that they have used to model a new situation in the \emph{real world}. The emphasis there is needed because ``Data Practices" is one of the key CT practices
yet in our opinion (as a critique of both ourselves and other coding-in-physics initiatives) this is not a strength of the content and tools that are currently in use. A possible exception is Tracker Video which gives the user some tools to simulate the dynamics of objects in videos \cite{TrackerVideo}, but it does not have a fully featured coding interface.


\begin{figure}
    \centering
    \includegraphics[width=2.5in]{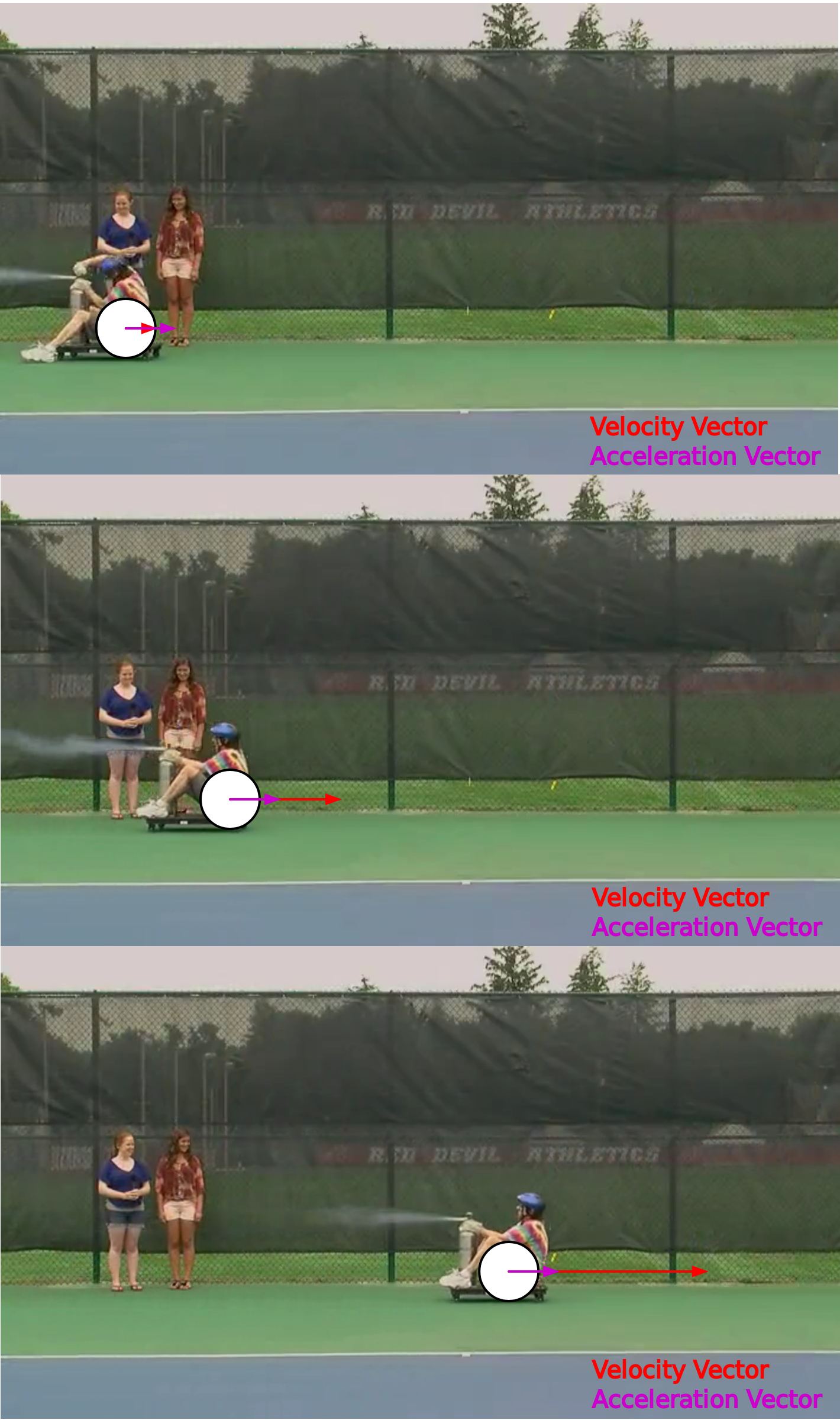}
    \caption{Screenshots at various times from a working demo (\url{http://go.osu.edu/workingdemo}) where our ``accelerate the blob" code from the STEMcoding hour of code activity is run simultaneously with a direct measurement video of a fire extinguisher cart experiment (constant acceleration process). This framework can easily be adapted to allow students to model projectile motion, for example, with code from the angry birds exercise. 
Video credit: Interactive Video Vignettes. }
    \label{fig:kart}
\end{figure}

To provide an example of what the future may hold, Fig.~\ref{fig:kart} highlights a proof-of-concept where a direct measurement video of a fire extinguisher cart experiment from Interactive Video Vignettes plays in the background and in the foreground there is a white circle that is an object being simulated. Students need to configure a code to have the correct physics (constant acceleration) and they adjust the initial velocity and acceleration to match the motion on the screen. A link to the demo is available at \url{http://go.osu.edu/workingdemo}.   It is not difficult to imagine similar activities for projectile motion or the ``coffee filter" experiment where students model the motion of a falling object with air drag. 
This is an exciting possibility because students can record slow motion videos of these experiments on modern smartphones and tablets with frame rates up to 240 frames per second. This is the same frame rate that many of excellent direct measurement videos from Peter Bohacek\footnote{These videos are now only accessible through Pivot Interactives \cite{pivotinteractives}} were recorded \cite{dmv}.  

\section{Conclusion}

More and more states are creating or adopting K-12 computer science standards \cite{k12CSstandards}, giving teachers more permission to integrate coding into physics courses than before. We hope this article provides helpful insights into the nature of CT, including ways that physics instruction already aligns with this instructional goal and ways that as a community we can work to help our students develop CT skills.











\acknowledgments

We acknowledge support from the 2017 AIP Meggers Award and the OSU Connect \& Collaborate grant program. We also thank the organizers of an NSF sponsored workshop on computational thinking that occurred May 2-5, 2019 in College Park, MD.

\bibliographystyle{unsrt}
\bibliography{main}
















\end{document}